\documentclass[sigconf, authorversion=true, nonacm=true]{acmart}

\usepackage{algpseudocode}
\usepackage{algorithm}
\usepackage{algorithmicx}
\algnewcommand\algorithmicforeach{\textbf{for each}}
\algdef{S}[FOR]{ForEach}[1]{\algorithmicforeach\ #1\ \algorithmicdo}
\usepackage{siunitx}
\usepackage{multirow}
\usepackage{xspace}
\algnewcommand{\algorithmicgoto}{\textbf{go to}}
\algnewcommand{\Goto}{\algorithmicgoto\xspace}
\algnewcommand{\Label}{\State\unskip}
\usepackage{bm}
\usepackage{physics} 
\AtBeginDocument{\RenewCommandCopy\qty\SI}

\begin{document}

\title[Whack-a-mole Online Learning]{Whack-a-mole Online Learning: Physics-Informed Neural Network for Intraday Implied Volatility Surface}

\author{Kentaro Hoshisashi}
\authornote{Corresponding author}
\email{k.hoshisashi@ucl.ac.uk}
\affiliation{%
  \department{Department of Computer Science}
  \institution{University College London}
  \city{London}
  \country{United Kingdom}
}
\author{Carolyn E. Phelan}
\email{carolyn.phelan.14@ucl.ac.uk}
\affiliation{%
  \department{Department of Computer Science}
  \institution{University College London}
  \city{London}
  \country{United Kingdom}
}
\author{Paolo Barucca}
\email{p.barucca@ucl.ac.uk}
\affiliation{%
  \department{Department of Computer Science}
  \institution{University College London}
  \city{London}
  \country{United Kingdom}
}

\renewcommand{\shortauthors}{Hoshisashi, et al.}

\begin{abstract}
Calibrating the time-dependent Implied Volatility Surface (IVS) using sparse market data is an essential challenge in computational finance, particularly for real-time applications. This task requires not only fitting market data but also satisfying a specified partial differential equation (PDE) and no-arbitrage conditions modelled by differential inequalities. This paper proposes a novel Physics-Informed Neural Networks (PINNs) approach called Whack-a-mole Online Learning (WamOL) to address this multi-objective optimisation problem. WamOL integrates self-adaptive and auto-balancing processes for each loss term, efficiently reweighting objective functions to ensure smooth surface fitting while adhering to PDE and no-arbitrage constraints and updating for intraday predictions. In our experiments, WamOL demonstrates superior performance in calibrating intraday IVS from uneven and sparse market data, effectively capturing the dynamic evolution of option prices and associated risk profiles. This approach offers an efficient solution for intraday IVS calibration, extending PINNs applications and providing a method for real-time financial modelling.
\end{abstract}



\maketitle

\section{Introduction}
The advent of deep learning has significantly transformed the modelling of complex systems, with artificial neural networks (ANNs) increasingly employed to parameterise unknown functions. Within this paradigm, physics-informed neural networks (PINNs) \cite{raissi2019physics} have emerged as an effective method for addressing forward and inverse problems involving partial differential equations (PDEs), incorporating both noisy data and physical laws in a learning framework. Applying this approach to real-world, real-time financial applications is challenging but potentially beneficial, particularly in the domain of option pricing and risk management.

Calibrating the time-dependent Implied Volatility Surface (IVS) is a crucial issue in quantitative finance that aligns well with the PINNs framework in terms of PDE-constrained problems. Previous research on IVS calibration (or interpolation) using ANNs has explored various approaches, including adapted global optimisation \cite{liu2019neural, choudhary2023funvol, cohen2023arbitrage}, advanced network models \cite{bergeron2022variational, cao2021option, cont2022simulation, ning2023arbitrage}, and constraint penalisation \cite{itkin2019deep, ackerer2020deep, choudhary2023funvol, cont2022simulation}. Some studies focus on the dynamics of IVS \cite{cont2002dynamics, balland2002deterministicimplied}, which are associated with PDEs and have applied PINNs to the Black-Scholes PDE \cite{bai2022application, dhiman2023physics}. However, approaches that fully account for all constraints required to model the dynamics of the options market have been limited.

To comprehensively model the IVS, calibration must simultaneously satisfy the governing PDE financial principles, including no-arbitrage conditions, and fit sparse, uneven, and increasing market data. This complex interplay of requirements leads to a challenging multi-objective optimisation problem, particularly when dealing with the temporal evolution of the IVS and the need for real-time updates. This challenge is especially relevant in today's fast-paced financial markets, where accurate and timely IVS calibration is essential for effective pricing strategies. To address these challenges, we propose a novel calibration algorithm called Whack-a-mole Online Learning (WamOL), which builds upon the PINNs framework by integrating self-adaptive and auto-balancing processes through layered optimisation to ensure alignment with PDE conditions and derivative inequalities. Furthermore, WamOL leverages online learning techniques to facilitate real-time intraday calibration, enabling the model to adapt swiftly to evolving market conditions.

Our contributions are summarised as follows:
\begin{itemize}
    \item We introduce the Whack-a-mole Online Learning (WamOL) algorithm, which combines self-adaptive and loss-balancing learning techniques to calibrate time-dependent IVS using PINNs. WamOL captures IVS characteristics while adhering to no-arbitrage and PDE conditions in real-time applications with increasing volumes of data.
    \item We demonstrate WamOL's effectiveness in improving interpolation and extrapolation of the IVS, including smooth and appropriate sensitivity profiles across time. The algorithm's self-adaptive configurations enhance the model's overall effectiveness in handling the temporal evolution of the volatility surface.
    \item WamOL provides an enhanced interpretation tool for time-dependent IVS calibration and risk profiles, using automatic differentiation to offer an efficient solution that builds on recent computational advancements in quantitative finance applications.
\end{itemize}
This study addresses the challenges of multi-objective training in a dynamic environment, providing a novel approach to accurately represent the time-dependent nature of options market dynamics while satisfying essential financial constraints.

The remainder of this paper is organised as follows: Section 2 discusses the IVS calibration problem and no-arbitrage constraints. Section 3 introduces multi-objective optimisation, including PINNs, Derivative-Constrained PINNs (DCPINNs), and online incremental learning. Section 4 details the proposed WamOL algorithm. Section 5 describes the neural network formulation used in our approach. Section 6 presents the algorithms, including the detailed WamOL algorithm. Section 7 outlines our experimental design, including network settings, training configurations, and the real-time sequential data used. Section 8 presents our results, including backtesting on real market data, risk sensitivity analysis, and an examination of loss scalability and computational efficiency. Finally, Section 9 concludes the paper and discusses future work.

\section{IVS Calibration}
\subsection{Constrained PDE Calibration Problem}
The Implied Volatility Surface (IVS) represents modelled values derived from European option prices. We are given a complete filtered probability space {\small$\left(\Omega, \mathcal{F},\left(\mathcal{F}_t\right)_{t \in [0,T]}, \mathbb{P}\right)$} where {\small$\mathbb{P}$} is an associated risk-neutral measure. The price of a European option $\Tilde{V}$ at time $t$ is
\begin{equation}
    \Tilde{V}_t = B_\tau\mathbb{E}\left[\left(\phi(S_T - k)\right)^+ \mid \mathcal{F}_t\right],\quad \phi = \begin{cases}1, & \text{call option}\\-1, & \text{put option}\end{cases}
\end{equation}
where $S_T$ is the underlying price at maturity $T$, $S_t$ is the current underlying price, $k$ is the strike price, $\tau$ is time to maturity from time $t$ with $\tau:=T-t$, and $B_\tau$ is the zero-coupon bond price associated with risk-free rate. We denote $F_\tau$ as the forward price at time $t$ for delivery of asset $S$ at time $T$. In \cite{black1973pricing, black1976pricing, gatheral2011volatility}, the Black-Scholes formula for the forward (undiscounted) value of the option price $V_t$ with modelled volatility $\sigma_{\rm{imp}}$ is redefined as,
\begin{equation}
\begin{gathered}
    V_t (F_\tau, k, \tau, \phi, \sigma_{\rm{imp}}) = \phi F_\tau N\left(\phi d_1\right) - \phi k N\left(\phi d_2\right),\\
    d_1 = \frac{\ln\left(F_\tau / k\right) + (\sigma_{\rm{imp}}^2 / 2) \tau}{\sigma_{\rm{imp}}\sqrt{\tau}}, \quad d_2 = d_1 - \sigma_{\rm{imp}}\sqrt{\tau},
    \label{eq: BS_formula}
\end{gathered}
\end{equation}
where $F_\tau = S_t / B_\tau$, $N(\cdot)$ is the cumulative normal distribution function.

To generalise for implied volatility in $k, \tau \in [0, \infty)$, \cite{dupire1994pricing} proposed the local volatility (LV) model, in which the option prices satisfy the PDE in forward system \cite{andreasen2011volatility} and PDE residuals $f$,
\begin{equation}
    f\left(k, \tau\right) := \pdv{V}{\tau} - \frac{1}{2} \sigma_{\rm{LV}}^2 \left(k, \tau\right) k^2 \pdv[2]{V}{k} = 0,
    \label{eq: LV_PDE}
\end{equation}
with initial condition $V_{t, \tau=0} = \left(\phi(F_0 - k)\right)^+$, and boundary conditions 
\begin{equation}     
    V_{t, k=\infty} = \begin{cases}0,&\phi=1\\\infty,&\phi=-1\end{cases},\quad
    V_{t, k=0} = \begin{cases}F_\tau,&\phi=1\\0,&\phi=-1\end{cases}.
    \label{eq: boundary_conditions}
\end{equation}
Plugging it into the formula in \eqref{eq: BS_formula}, one obtains a conversion function between $\sigma_{\rm{imp}}$ and $\sigma_{\rm{LV}}$ with respect to $k$ and $\tau$ with $w:={\sigma_{\rm imp}}^2\tau$ and log scaled strike $\widehat{k}=\ln\left(k/F_\tau\right)$ in \cite{gatheral2011volatility},
\begin{equation}
    \sigma_{\rm LV}^2 =\frac{\pdv{w}{\tau}}{1-\frac{\widehat{k}}{w} \pdv{w}{\widehat{k}}+\frac{1}{4}\left(-\frac{1}{4}-\frac{1}{w}+\frac{\widehat{k}^2}{w^2}\right)\left(\pdv{w}{\widehat{k}}\right)^2+\frac{1}{2} \pdv[2]{w}{\widehat{k}}}
    \label{eq: LV_vol_from_BS_vol}
\end{equation}
to be fit with the PDE \eqref{eq: LV_PDE}.
In this study, we define the IVS calibration as identifying the multivariate function respecting the prices $\Phi\left(x\right)$, associated with IVS as a function $\varphi\left(x\right) \geq 0$ with inputs $x:=(\widehat{k}, \tau, t)$,
\begin{equation}
    \Phi(x) = V_t \left(F_\tau, k, \tau, \phi, \varphi\left(x\right)\right).
    \label{eq: price_function_by_Phi}
\end{equation}
The inverse problem of the IVS is that, given limited option prices, we would like to identify the function of IVS with respect to $x$, redefined as $\sigma_{\rm{imp}}\left(x\right)$, to fit the premium resulting from \eqref{eq: BS_formula} and also satisfy the PDE. Based on \eqref{eq: BS_formula}--\eqref{eq: LV_vol_from_BS_vol}, once $\varphi$ is determined, we can analytically obtain the option price $\widehat{\Phi}:=B_\tau\Phi$. Furthermore, $\Phi$ is second differentiable whenever $\varphi$ is second differentiable, allowing the representation of the PDE in \eqref{eq: LV_PDE}.

\subsection{No-Arbitrage Constraints}
A key challenge in this problem is the scarcity of option price data. To address this, one approach is to incorporate additional relevant information, including no-arbitrage conditions. Option price calibration is constrained by sparse data and must adhere to no-arbitrage principles. These principles state that market prices should not allow for guaranteed returns exceeding the risk-free rate. We consider the necessary and sufficient conditions for no-arbitrage in \cite{carr2005note, ait2003nonparametric}, which are represented as non-strict inequalities involving several first and second derivatives,
\begin{equation}
    -1 \leq \phi\pdv{V}{k}\leq 0,\;\;\; \pdv[2]{V}{k} \geq 0,\;\;\;  \pdv{V}{\tau}\geq 0.
    \label{eq: no-arbitrage conditions}
\end{equation}
From the above, no-arbitrage conditions require these derivatives to have a specific sign. The standard architecture does not automatically satisfy these conditions when calibrating with a loss function simply based on the mean squared error (MSE) for the prices. These no-arbitrage conditions are important as they ensure that the calibrated IVS is consistent with fundamental principles of financial economics, preventing unrealistic or exploitable pricing scenarios.

\section{Multi-Objective Optimisation}
IVS calibration is a complex problem that requires fitting a model to PDE and no-arbitrage conditions, which impose additional constraints on the derivative values of the function surface. Forming a new class of data-efficient universal function approximations, we would apply deep neural networks, which are increasingly being applied to model and simulate such systems with PDEs.

\subsection{Physics-Informed Neural Networks (PINNs)}
Physics-Informed Neural Networks (PINNs), introduced in \cite{raissi2019physics}, is a deep learning framework that incorporates underlying physical laws into the architecture through PDEs and solves this PDE system as an optimisation problem by minimising the total loss:
\begin{equation}
    \mathcal{L} := \mathcal{L}_0 + \mathcal{L}_b + \mathcal{L}_f,
\end{equation}
\begin{equation}
\begin{gathered}
    \mathcal{L}_0 := \frac{1}{N_0} \sum_{i=1}^{N_0} \left| \widehat{\Phi}(x^{(i)}) - \widehat{V}^{(i)} \right|^2,\\
    \mathcal{L}_b := \frac{1}{N_b} \sum_{i=1}^{N_b} \left| \mathcal{B}[\Phi(x^{(i)})] \right|^2,\\
    \mathcal{L}_f := \frac{1}{N_f} \sum_{i=1}^{N_f} \left|f[\Phi(x^{(i)})] \right|^2,
\label{eq: pinns_metrics}
\end{gathered}
\end{equation}
where $\mathcal{L}_0$ represents the mean squared error between the model predictions $\widehat{\Phi}(x^{(i)})$ and observed values $\widehat{V}^{(i)}$, $\mathcal{L}_b$ enforces the boundary conditions, and $\mathcal{L}_f$ enforces the physical constraints imposed by the PDE at a finite set of collocation points. The operators $\boldsymbol{f}$ and $\boldsymbol{\mathcal{B}}$ represent the PDE residuals and boundary conditions, respectively, both acting on $\Phi$. $N_0$, $N_b$, and $N_f$ denote the number of data points, boundary points, and collocation points, respectively, while $x^{(i)}$ represents the $i$-th point in each corresponding set.

\subsection{Derivative-Constrained PINNs (DCPINNs)}
\label{subsection: Loss Considerations for Inequality-Differential Constraints}
We now consider the generalisation of PINNs with partial derivative inequality conditions, called Derivative-Constrained PINNs (DCPINNs). We assume
\begin{equation}
    h[\Phi(x)], x \in \Omega
\end{equation}
where $h[\cdot]$ represents a set of differential operators acting on an inequality equation in the problem, e.g., differential terms in \eqref{eq: no-arbitrage conditions}. To fit with inequality constraints, a loss $\mathcal{L}_h$ to be minimised is defined as,
\begin{equation}
    \mathcal{L}_h := \frac{1}{N_h} \sum_{i=1}^{N_h} \gamma \circ \left|h\left(\Phi(x^{(i)})\right)\right|^2.
\label{eq: dcpinns_metric}
\end{equation}
\begin{equation}
    \gamma(x) = \begin{cases}{x}, & \text{if inequality is not satisfied} \\ 0, & \text{otherwise}\end{cases}
\end{equation}
where $\mathcal{L}_h$ enforces the structure imposed by inequality with differential operators at a finite set of collocation points, $\circ$ is the operation of the applied nonlinear function, and $\gamma$ reflects the loss whether the inequality is satisfied or not.

\subsection{Online Incremental Learning}
Online incremental learning is a machine learning paradigm in which the model is continuously updated as new data becomes available, making it particularly essential for scenarios with streaming data. The process from $t_1$ to $t_2$ can be represented as:
\begin{equation}
\theta_{t_2} = \theta_{t_1} + \eta \Delta\theta_{t_1}
\end{equation}
where $\theta_{t_1}$ represents the set of model parameters at time $t$, $\theta_{t_2}$ represents the updated parameters at time $t_2$, $\eta$ is the learning rate, and $\Delta\theta_{t_1}$ is the update to the parameters based on the new data points.

The update term $\Delta\theta_{t_1}$ involves the gradient of a loss function with respect to the current parameters and the new data point(s)
\begin{equation}
\Delta\theta_{t_1} = -\nabla \mathcal{L}(\theta_{t_1}, x_{t_2}, \widehat{V}_{t_2}),
\end{equation}
where $\nabla \mathcal{L}$ is the gradient of the loss function $\mathcal{L}$, $x_t$ is the set of input feature vector at time $t$, and $\widehat{V}_t$ is the set of observable outputs at time $t$. We then can define the updated loss as
\begin{equation}
    \mathcal{L}_t := \frac{1}{N_t} \sum_{i=1}^{N_t} \zeta \circ \left| \widehat{\Phi}(x_t^{(i)}) - \widehat{V}_t^{(i)} \right|^2,
\end{equation}
where
\begin{equation}
    x_t:=\left\{x^{(i)} \mid t^{(i)}<t\right\}, \widehat{V}_t:=\left\{\widehat{V}^{(i)} \mid t^{(i)}<t\right\},
\end{equation}
and function $\zeta$ controls the weights for increased data importance and information loss due to time decay. This approach is particularly relevant for IVS calibration in real-world financial markets, where new option price data is constantly streaming in, and market conditions can change rapidly. Other research applies different types of control policies that naturally balance the exploration-exploitation trade-off throughout the online phase, such as in \cite{waldon2024dare}.

\section{Whack-a-mole Online Learning}
\label{section: Whack-a-mole Online Learning (WamOL)}
IVS calibration aims to identify the surface $\Phi_{\theta}(x)$ by finding the optimal parameter set $\theta$ with inputs $x:=(\widehat{k}, \tau, t)$. The standard architecture does not automatically satisfy all PDEs and no-arbitrage conditions across temporal dimensions. Here, we consider optimisation problems that include multiple losses, not only for the surface but also for its derivatives, following \cite{EasyChair:15219}. The total cost to be minimised is,
\begin{equation}
    \widehat{\mathcal{L}} := \underbrace{\lambda_t \widehat{\mathcal{L}}_t}_{\text{Increasing Inputs}} + \lambda_b \widehat{\mathcal{L}}_b + \underbrace{\lambda_f \widehat{\mathcal{L}}_f}_{\text{PDE residuals}} + \underbrace{\Sigma_\alpha \lambda_{h_\alpha} \widehat{\mathcal{L}}_{h_\alpha}}_{\text{No-arbitrage penalties}},
    \label{eq: total loss function in WamOL}
\end{equation}
\begin{equation}
    \widehat{\mathcal{L}}_t :=\frac{1}{N_t} \sum_{i=1}^{N_t} m_t^{(i)} \zeta \circ \left|\widehat{\Phi}_{\theta}\left(x_t^{(i)}\right)-\widehat{V}_t^{(i)}\right|^2,
    \label{eq: WamOL price loss}
\end{equation}
\begin{equation}
    \widehat{\mathcal{L}}_b :=\frac{1}{N_b} \sum_{i=1}^{N_b} m_b^{(i)} \left|\Phi_{\theta}\left(x_b^{(i)}\right)-V_b^{(i)}\right|^2,
\end{equation}
\begin{equation}
    \widehat{\mathcal{L}}_f := \frac{1}{N_f} \sum_{i=1}^{N_f} m_f^{(i)} \left|f\left(\Phi_{\theta}\left(x^{(i)}\right)\right)\right|^2,
\end{equation}
\begin{equation}
    \widehat{\mathcal{L}}_{h_\alpha} := \frac{1}{N_{h_\alpha}} \sum_{i=1}^{N_{h_\alpha}} m_h^{(i)} \left| \gamma_\alpha \circ {h_\alpha}\left(\Phi_{\theta}\left(x^{(i)}\right)\right) \right|^2.
    \label{eq: WamOL no-arbitrage loss}
\end{equation}
where $V_t^{(i)}$ is the observed premium increasing by time, $i=1,\ldots,N_t$ from the dataset, and $\widehat{\mathcal{L}}_{h_\alpha}$ represents each penalty term corresponding to \eqref{eq: no-arbitrage conditions}, i.e., $\mathcal{L}_{h_{\rm k}}$, $\mathcal{L}_{h_{\rm kk}}$ and $\mathcal{L}_{h_{\rm \tau}}$ correspond to the differential terms in \eqref{eq: no-arbitrage conditions}, respectively. The modifications from standard PINNs configurations are the multiplier $\lambda$ for each loss category and the weights $m^{(i)}$ for individual data points within each loss term.

\label{Whack-a-mole Online Learning (WamOL)}
\begin{figure*}[htbp]
    \centering
    \includegraphics[width=\textwidth]{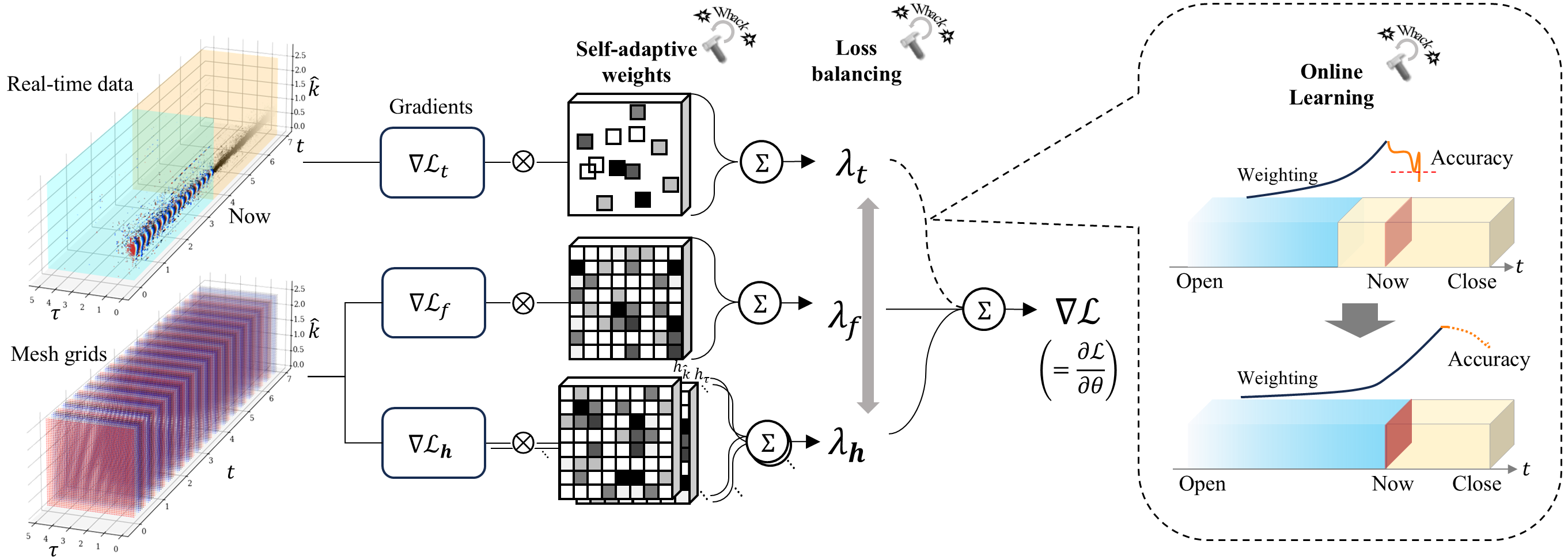}
    \caption{The architecture of Whack-a-mole Online Learning (WamOL), illustrating its three key components: (1) Self-adaptive weights for gradient intensification, (2) Loss balancing to address multi-scale imbalances, and (3) Online learning with time decay for adaptive model updates.}
    \label{Figure: WamOL_Architecture}
    \Description[WamOL architecture.]{The whole architecture of Whack-a-mole Online Learning (WamOL).}
\end{figure*}
The concept of Whack-a-mole Online Learning (WamOL), proposed in this study, involves a combination of three "whacks" in the learning process. The first "whack" intensifies the gradient of individual losses in categorised losses to enhance local constraints, especially the objective for inequality derivative constraints. The second "whack" addresses the multi-scale imbalance between categorised losses in (\ref{eq: WamOL price loss}$\sim$\ref{eq: WamOL no-arbitrage loss}), particularly to mitigate the fluctuation of gradient values from epoch to epoch due to the inequality feature in \eqref{eq: WamOL no-arbitrage loss}. The third "whack" controls online incremental learning, which restarts the learning process corresponding to the increased inputs by time decay and decreased accuracy.

\paragraph{First whack: self-adaptive weights}
In alignment with the neural network philosophy of self-adaptation, WamOL applies a straightforward procedure with fully trainable weights to generate multiplicative soft-weighting and attention mechanisms, inspired by \cite{mcclenny2020self, wang2023expert}. The first "whack" proposes self-adaptive weighting that updates the loss function weights via gradient ascent concurrently with the network weights. We minimise the total cost with respect to $\theta$ but also maximise it with respect to the self-adaptation weight vectors $m$ at the $p$-th epoch:
\begin{equation}
    m_\beta^{(j)}(p+1) = m_\beta^{(j)}(p) + \eta_m\nabla_{m_\beta^{(j)}} \widehat{\mathcal{L}}_\beta(p),
\end{equation}
where $\beta$ specifies each loss $\beta \in \left\{0,b,f,h_{\rm k},h_{\rm kk},h_{\rm \tau}\right\}$ and $\eta_m$ is the learning parameter.
In the learning step, the derivatives with respect to self-adaptation weights are increased when the constraints of derivative terms are violated.

\paragraph{Second whack: loss balancing}
For the second "whack", loss-balancing employs the following weighting function in the loss function based on Eq. \eqref{eq: total loss function in WamOL}. Considering updates at the $p$-th epoch:
\begin{equation}
    \lambda_\beta(p+1) = \begin{cases}{1}, & \text{if $\overline{\abs{\nabla_\theta\widehat{\mathcal{L}}_\beta(p)}}=0$} \\ \frac{1}{2}\left(\lambda_\beta(p) + \frac{\sum_\beta \overline{\abs{\nabla_\theta\widehat{\mathcal{L}}_\beta(p)}}}{\overline{\abs{\nabla_\theta\widehat{\mathcal{L}}_\beta(p)}}}\right), & \text{otherwise} \end{cases}
\end{equation}
where $\nabla_i$ is the partial derivative vector (gradient) with respect to the $i$-th input vector (value), and $\overline{\abs{\cdot}}$ indicates the average of the absolute values of the elements in the vector. The choice of absolute values, instead of the squared values in \cite{wang2023expert}, is to avoid overlooking outlier gradients from the individual losses, as most elements of $\nabla\mathcal{L}_h$ are zero values in the final stages of training.

\paragraph{Third whack: online learning}
To maintain the balance between adaptivity and stability, this study introduces a weighting function for controlling the weights for increased data importance and information loss due to time decay. We apply:
\begin{equation}
\zeta(t_i, t) = e^{-\mu (t-t_i)}
\end{equation}
in Eq. \eqref{eq: WamOL price loss}. Thus, the updated price loss from time $t_1$ to $t_2$ with streamed data is:
\begin{equation}
    \widehat{\mathcal{L}}_{t_2}\left(x_{t_2}\right) := \frac{N_{t_1}}{N_{t_2}}\zeta\left(t_1, t_2\right) \widehat{\mathcal{L}}_{t_1}\left(x_{t_1}\right) + \frac{N_{t_2} - N_{t_1}}{N_{t_2}} \widehat{\mathcal{L}}_{\Delta t}\left(x_{t_2} \setminus x_{t_1}\right),
\end{equation}
where $x_t$ represents the grids of updated observed variables, i.e., European options, $\widehat{\mathcal{L}}_{\Delta t}$ represents the price error between $t_1$ and $t_2$, and $\mu$ is a predefined coefficient.

The proposed method automatically adjusts the weights of loss terms based on their relative magnitudes during training at user-specified intervals throughout the training process. Unlike standard online learning approaches that typically focus on minimising a single loss function, WamOL dynamically balances multiple objectives. This approach allows for more effective handling of the complex, multi-objective nature of IVS calibration in real-time market environments.

\section{Neural Network Formulation}
\label{section: Neural Network Formulation}
In WamOL, we apply a simple but deep feed-forward neural network architecture, a multilayer perceptron (MLP). Let $L \geq 2$ be an integer representing the depth of the network; we consider an MLP with input vector $x \in \mathbb{R}^n$, $L$ hidden layers, and one output value $y \in \mathbb{R}$. We can consider the price function $\Phi$ in \eqref{eq: price_function_by_Phi} which includes MLP as a multivariate function $\varphi$ depending on the variables $x$, i.e., $\varphi: \mathbb{R}^3 \rightarrow \mathbb{R}$,
\begin{equation}
\begin{gathered}
    \varphi\left(x\right) = \varsigma \circ A_L \circ f_{L-1} \circ A_{L-1} \circ \cdots \circ f_1 \circ A_1\left(x\right),
\end{gathered}    
\end{equation}
where for $l=1,\ldots , L$, $A_{l}: \mathbb{R}^{d_{l-1}} \rightarrow \mathbb{R}^{d_{l}}$ are affine functions as $A_l(x_{l-1}) = W_l^\mathsf{T} x_{l-1} + b_l$, and $d_l$ is the number of neurons in the layer $l$ for $x_{l-1} \in \mathbb{R}^{d_{l-1}}$, with $W_l \in \mathbb{R}^{d_{l-1} \times d_l}$ and $b_l \in \mathbb{R}^{d_l}$,  $d_{0}=n$, $d_{L}=1$, and $x_t=x$. $f_l$ and $\varsigma$ are activation functions which are applied component-wise. To restrict the positivity of implied volatility and enforce the boundary conditions \eqref{eq: boundary_conditions}, as stated in \cite{lee2005implied}, we applied the softplus function as the last activation function $\varsigma\left(x\right) = {\rm ln}\left(1+e^x\right) > 0$. Given a dataset $x$ and a loss function $\mathcal{L}(x, \varphi)$, the model $\varphi$ is found by fitting the parameters $\theta$ (i.e., $\textbf{W}$ and $\textbf{b}$) which minimise the loss function.

A challenge lies in that Eq.~\eqref{eq: total loss function in WamOL} involves derivatives with respect to $x$, which are also functions of the parameters. When numerical approximation of derivatives is used, it could result in slow or inaccurate solutions. To solve this, this study utilises an extended backpropagation algorithm from \cite{rumelhart1986learning} with Automatic Differentiation for the gradient in Eq.~\eqref{eq: total loss function in WamOL} through exact derivative formulations, which require activation functions in the network to be second-order differentiable or higher. It is noted that non-differentiable functions, e.g., ReLU or ELU, need additional consideration at singular points. If all activation functions are second-order differentiable, the same is true for the whole network as shown in \cite{hornik1990universal, hoshisashi2023no}.

\section{Algorithms}
\label{section: Algorithms}
This section introduces Algorithms of the WamOL for multi-objective problems as an expansion of the work in \cite{raissi2019physics}, which controls the inequality loss of the partial derivatives of a neural network function with respect to its input features.
\begin{algorithm}
\caption{Whack-a-mole Online Learning (WamOL)}
\label{alg: WamOL}
\begin{algorithmic}
\Require \hspace{-5pt} $\left\{\left(x_{t_1}, V_{t_1}\right)^{(i)}\right\}_{i=1}^{N_{t_1}}$, $\left\{\left(x_{t_2}, V_{t_2}\right)^{(i)}\right\}_{i=1}^{N_{t_2}}$, $\left\{x_f^{(i)} \right\}_{i=1}^{N_f}$, $\left\{x_\textbf{h}^{(i)} \right\}_{i=1}^{N_\textbf{h}}$ 
\Ensure Neural network parameters $\theta$
\vspace{3pt}\State Consider a deep NN $\varphi_\theta(x)$ with $\theta$, and a loss function 
$$\widehat{\mathcal{L}}:= \sum_\beta \lambda_{\beta} \widehat{\mathcal{L}}_{\beta}\left(m_\beta, x_\beta \left(, V_\beta\right)\right),$$
where $\widehat{\mathcal{L}}_{\beta}$ denotes the categorised loss with $\beta \in \left\{t_1,f,\textbf{h}\right\}$, $m_\beta=\textbf{1}$ are weight vectors for individual loss and $\lambda_\beta=1$ are dynamic parameters to balance between the different categorised loss.
\vspace{3pt}\State Update $\theta$ with pre-determined parameters $\eta$, $\eta_m$, $q_m$, $q_\lambda$, $p_{\rm max}$:
\For{$p=1, \ldots, p_{\rm max}$}
    \State Compute $\nabla_{\theta} \widehat{\mathcal{L}}_\beta(p)$ by automatic differentiation
    \Label \;\;\;\;  \texttt{marker:}
    \If{$p \equiv 0 \pmod{q_m}$}
        \State \textit{Self-adaptive weights:} Update $m_\beta$ by$$m_\beta^{(j)}(p+1) = m_\beta^{(j)}(p) + \eta_m \nabla_{m_\beta^{(j)}} \widehat{\mathcal{L}}_\beta(p),$$
        \State where $m_\beta(p), \widehat{\mathcal{L}}_\beta(p)$ shows at $p$-th iteration.
    \EndIf
    \If{$p\equiv 0 \pmod{q_\lambda}$}
        \State \textit{Loss balancing:} Update $\lambda_\beta$ by {\small $$\hspace{28pt}\lambda_\beta(p+1) = \begin{cases}{1}, & \text{\hspace{-36pt}if $\overline{\abs{\nabla_\theta\widehat{\mathcal{L}}_\beta(p)}}=0$} \\ \frac{1}{2}\left(\lambda_\beta(p) + \frac{\sum_\beta \overline{\abs{\nabla_\theta\widehat{\mathcal{L}}_\beta(p)}}}{\overline{\abs{\nabla_\theta\widehat{\mathcal{L}}_\beta(p)}}}\right), & \text{otherwise} \end{cases}$$}
        \State where $\overline{\abs{\cdot}}$ is mean of the absolute values of elements.    
    \EndIf
    \State Update the parameters $\theta$ via gradient descent, e.g.,
    \State \hspace{15pt}$\theta(p+1)=\theta(p)-\eta \nabla_{\theta} \widehat{\mathcal{L}}\left(p\right)$.
\EndFor
\State \textit{Online learning:} Update $\theta$ when the accuracy becomes worse:
\If{$\mathcal{L}_{\Delta t}\left(x_{t_2} \setminus x_{t_1}\right) > \varepsilon$}
    {\small $$\;\;\;\;\;\;\widehat{\mathcal{L}}_{t_2}\left(x_{t_2}\right) := \frac{N_{t_1}}{N_{t_2}}\zeta\left(t_1, t_2\right) \widehat{\mathcal{L}}_{t_1}\left(x_{t_1}\right) + \frac{N_{t_2} - N_{t_1}}{N_{t_2}} \widehat{\mathcal{L}}_{\Delta t}\left(x_{t_2} \setminus x_{t_1}\right).$$}
    \State $p \leftarrow 1$ \\
    \;\;\;\; \Goto \texttt{marker}
\EndIf
\State \textbf{Return} $\theta$
\end{algorithmic}
\end{algorithm}

Algorithm \ref{alg: WamOL} exhibits WamOL characteristics that distinguish it from conventional learning methods. First, the computation points $x_{\{f,h\}}$ for the derivatives of the MLP do not correspond to the points of the training dataset $x$, as the latter is typically sparse and unevenly distributed. The algorithm adjusts the derivatives to fit mesh grids, thus capturing derivative data across a wide array of input features. Secondly, the objective function $\mathcal{L}$ depends not only on the MLP's direct output but also on its derivatives, as specified in Eq.~\eqref{eq: total loss function in WamOL}, all of which depend on identical network parameters. Lastly, WamOL facilitates control over the weights for increased data importance and information loss due to time decay and continuously updates parameters as new data becomes available, making it particularly suitable for scenarios with streaming data. WamOL enables balancing accurate calibration and categorised losses, which consist of PDE residuals and derivative inequalities in real-time updated situations, as described in Section \ref{section: Whack-a-mole Online Learning (WamOL)}.

\section{Experimental Design}
\label{section: Experimental Design}
\subsection{Network Setting and Training Configuration}
\label{Neural Network Setting and Training Configuration}
In selecting appropriate network architectures and adopting suitable learning algorithms, we refer to approaches and configurations in \cite{wang2023expert, lu2021physics} for specific measures to improve learning efficiency and accuracy. In our experiment, the network architecture $\varphi$ is a deep setting with four hidden layers ($L=5$), each containing 64 neurons and a hyperbolic tangent activation function for smooth activations. Additionally, we apply non-dimensionalisation for input as log-moneyness, $\widehat{k}=\ln\left(k/F_\tau\right)$, and Glorot initialisation for $\theta$. Reference grids for PDE residuals ($N_f$) and no-arbitrage constraints ($N_{h_{\left\{\rm{k}, \rm{kk}, \rm{\tau}\right\}}}$) are defined as dense rectangular mesh grids, which are equally distributed with $21$ points in $\widehat{k}\in[-3.0, 1.2]$, $\tau\in[0, 5]$, and $t\in[0, 6.75]$ (hours) in training. In evaluation, we apply a dense grid, i.e., ($31$, $31$, $136$) grids for ($\widehat{k}$, $\tau$, $t$) to assess the interpolation ability. The hyperparameters used are $\mu=0.5$, $\eta_m=1$, $q_m = q_\lambda = 100$, and $p_{\rm max}=5,000$. We employ Adam optimisation \citep{kingma2014adam} with weight decay, starting with a learning rate of $\eta=10^{-3}$ and an exponential decay rate of $0.9$ for every $1,000$ steps.
The models compared in this study, omitting $\mathcal{L}_b$ as boundary conditions are always satisfied by $\varphi \geq 0$, are:
\begin{equation}
\begin{aligned}
\text{MLP}& \;\; \mathcal{L} := \mathcal{L}_t \\
\text{PINNs}& \;\; \mathcal{L} := \mathcal{L}_t + \mathcal{L}_f \\
\text{WamOL}& \;\; \widehat{\mathcal{L}} := \lambda_t \widehat{\mathcal{L}}_t + \lambda_f \widehat{\mathcal{L}}_f + \lambda_{h_{\rm k}} \widehat{\mathcal{L}}_{h_{\rm k}} + \lambda_{h_{\rm kk}} \widehat{\mathcal{L}}_{h_{\rm kk}} + \lambda_{h_{\rm \tau}} \widehat{\mathcal{L}}_{h_{\rm \tau}}
\end{aligned}
\end{equation}
For computation, we use differentiable solvers developed in JAX/Flax \cite{jax2018github, flax2020github}, which are suitable for inverse problems when modelling an unknown field. The experiments are conducted using Google Colab\footnote{Google Colab. \url{https://colab.research.google.com}}, which offers GPU computing on the NVIDIA Tesla T4 with $15$ GB of video random access memory. The results reported in the figures represent the values with synchronised seed values for random variables across the compared models.

\subsection{Real-time Sequential Data}
\label{subsection: Real Data}
We conducted backtesting on extensive intraday options data for WamOL using a dataset of S\&P 500 options. The historical data comprised intraday traded prices of S\&P 500 options for 248 business days from October 1, 2022, to September 30, 2023. This dataset contains approximately $500,000$ data points per day with $1$ ms precision, obtained from CBOE DataShop\footnote{CBOE DataShop, (2023). \url{https://datashop.cboe.com/option-trades}}. In the experiment, we aggregated the data into one-minute intervals ($\Delta t \approx 0.017$) to reduce the influence of outliers. The data source provided spot prices, selecting out-of-the-money (OTM) options, and we derived forward prices using linear regression based on put-call parity. Each trading day, the regular market hours were from 9:30 AM to 4:15 PM ET.

\begin{table}[htbp]
\begin{center}
\caption{Statistics of intraday prices of S\&P 500 options for 248 business days.}
\begin{tabular}{@{}lrrl@{}}
    \toprule
        Elements [per regular hours in a day] & Mean & \scriptsize{(min.)} & \scriptsize{(max.)} \\
    \midrule
        No. of traded prices & 435,193 & \scriptsize{(148,126)} & \scriptsize{(599,658)} \\
        No. of unique grids $(\hat{k}, \tau)$ & 5,002 & \scriptsize{(3,637)} & \scriptsize{(6,113)} \\
        Pct. of call options & 48.6\%  & \scriptsize{(42.5\%)} & \scriptsize{(53.7\%)} \\
        Pct. of short-term ($\tau \leq 0.8\dot{3} (1 \text{ month})$) & 94.4\% & \scriptsize{(90.4\%)} & \scriptsize{(96.7\%)} \\
        Pct. of near ATM ({\small $\widehat{k} \in [-0.1, 0.1]$}) & 97.1\% & \scriptsize{(94.8\%)} & \scriptsize{(98.3\%)} \\
    \bottomrule
\end{tabular}
\label{table: SP500_options_cboe}
\end{center}
\end{table}
\begin{figure}[htbp]
    \centering
    \includegraphics[width=0.9\columnwidth]{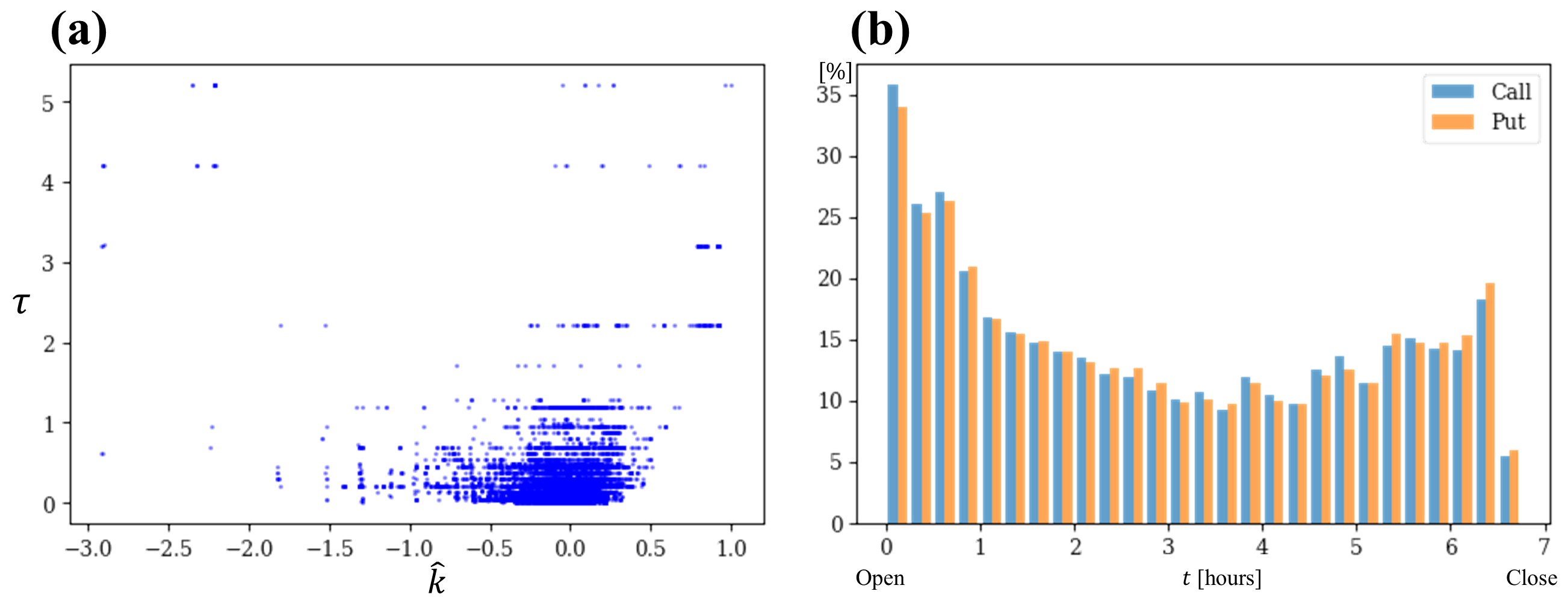}
    \caption{Intraday market distribution. (a) Distributed grids of traded options intraday. (b) The histogram shows the percentage of traded options by the time of the day.}
    \label{fig: wamol_market_dist}
    \Description[Intraday market distribution.]{Intraday market distribution. (a) Distributed grids of traded options intraday. (b) The histogram shows the percentage of traded options by the time of the day.}
\end{figure}
Table \ref{table: SP500_options_cboe} and Figure \ref{fig: wamol_market_dist} demonstrate that the distribution of input variables for intraday traded options is very uneven. The vast majority of trades (over $90\%$) involve options with short expiration dates of less than one month. Furthermore, about $97\%$ of these trades are concentrated near the at-the-money strike. Compared to typical scenarios using mesh grid data, this uneven distribution presents a more challenging task for calibrating IVS.

\section{Results}
\label{section: Results}
\subsection{Backtesting on Real Market Data}
\label{subsection: Backtesting on Real Market Data}
The backtesting procedure involved calibrating the IVS using WamOL and the baseline models for each trading day as presented in Section \ref{subsection: Real Data}. The calibrated surfaces were then used to assess the accuracy of option price predictions, PDE residual losses, and adherence to no-arbitrage constraints.

\begin{figure}[htbp]
\centering
\includegraphics[width=\columnwidth]{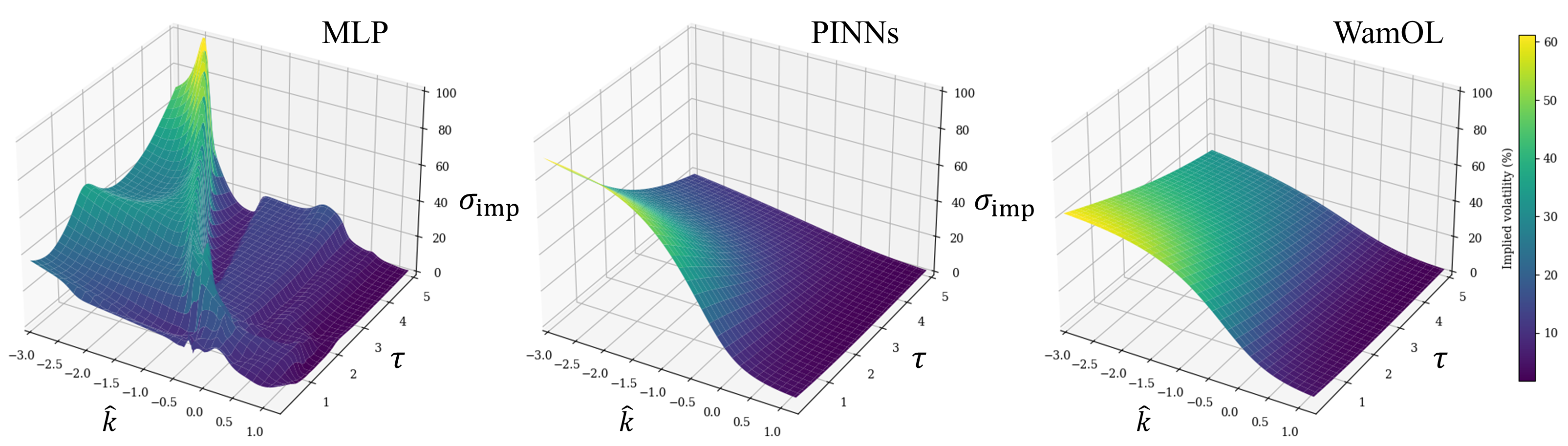}
\caption{Predicted surfaces of IVS at $t = 6.75$ based on trained data for $t<5$, exemplified by data on October 14, 2022.}
\label{fig: wamol_surfaces}
\Description[Predicted surfaces.]{Predicted IVS at $t = 6.75$ based on trained data for $t \leq 5$ on October 14, 2022.}
\end{figure}
Figure \ref{fig: wamol_surfaces} presents a typical visual comparison of these models' IVS predictions with respect to strike price and time to maturity, for a specific time point (close time, i.e., $t = 6.75$) based on training with intraday data for $t\leq5$, exemplified by data on October 14, 2022. The MLP model produces a rough surface, which appears unsuitable for accurate prediction. In contrast, both the PINNs and WamOL generate smooth surfaces, exhibiting high volatility in the short term and low volatility in the long term. WamOL demonstrates a consistent pattern of gradients in the strike direction across different maturities, likely a result of the imposed derivative constraints in the loss formulation.

\begin{table}[H]
    \centering
    \caption{Training and prediction errors on a logarithmic scale (mean) by models on real data. \textbf{Bold values} indicate lower (better) values.}
    \label{tab: backtesting_metrics}
    \begin{tabular}{clccccc}
    \toprule
    \multicolumn{2}{c}{Models} & $\mathcal{L}_t$ & $\mathcal{L}_f$ & $\mathcal{L}_{h_{\rm k}}$ & $\mathcal{L}_{h_{\rm kk}}$ & $\mathcal{L}_{h_{\rm \tau}}$\\
    \midrule
    \multirow{3}{*}{Training}
    &MLP & $\textbf{-5.41}$ & $-0.14$ & $-1.91$ & $-1.23$ & $-4.99$ \\
    &PINNs & $-4.63$ & $\textbf{-7.18}$ & $-2.71$ & $-9.82$ & $-10.1$ \\
    &WamOL & $-4.42$ & $-6.55$ & $\textbf{-4.48}$ & $\textbf{-14.0}$ & $\textbf{-12.1}$ \\
    \hline
    \multirow{3}{*}{Prediction}
    &MLP & $-4.56$ & $1.93$ & $0.52$ & $7.25$ & $-3.82$ \\
    &PINNs & $-4.81$ & $2.47$ & $-\textbf{inf}$ & $-1.88$ & $-9.82$ \\
    &WamOL & $\textbf{-4.86}$ & $\textbf{0.87}$ & $-\textbf{inf}$ & $\textbf{-2.00}$ & $\textbf{-12.36}$ \\
    \bottomrule
    \end{tabular}
\end{table}
Table \ref{tab: backtesting_metrics} presents key performance metrics averaged over the entire backtesting period for the models during both training ($t\leq3$) and prediction phases ($t>3$) averaged over $248$ days. The metrics are presented on a logarithmic scale, e.g., zero value shows $-\text{inf}$, where lower values indicate better performance. Each metric corresponds to the loss related to Eqs. \eqref{eq: pinns_metrics} and \eqref{eq: dcpinns_metric}. During training, MLP achieves the lowest $\mathcal{L}_t$, indicating the best fit to market prices, but performs poorly on other metrics. PINNs achieves the lowest PDE residual losses ($\mathcal{L}_f$). WamOL demonstrates the best performance in satisfying no-arbitrage constraints, with the lowest losses for no-arbitrage constraints. In the prediction phase, WamOL shows the best overall performance with the lowest $\mathcal{L}_t$, indicating better generalisation to new market data. PINNs achieves the lowest $\mathcal{L}_f$, but WamOL is close behind. Both PINNs and WamOL perfectly satisfy the delta constraint, while WamOL outperforms on the whole.

\begin{figure}[htbp]
    \centering
    \includegraphics[width=\columnwidth]{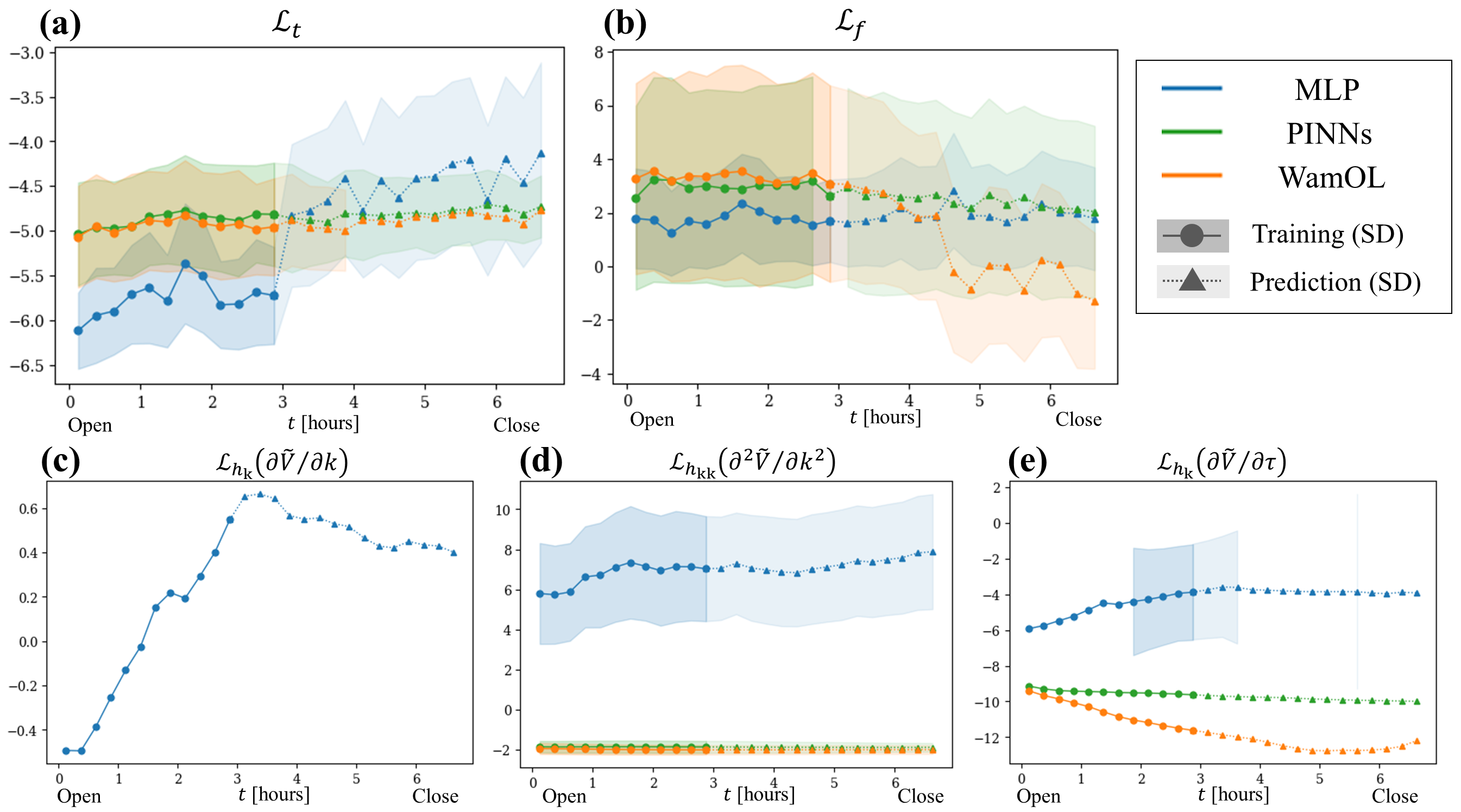}
    \caption{Backtesting results showing calibration and validation errors intraday on a logarithmic scale.}
    \label{fig: backtesting_results}
    \Description[Backtesting results.]{Backtesting results showing calibration and validation errors intraday.}
\end{figure}
Figure \ref{fig: backtesting_results} illustrates the performance of WamOL and the baseline models over the backtesting period intraday. Each error is shown as a function of training duration from market opening time to calibration time (solid lines) and validated using the calibrated model on data from the rest of the day (dotted lines) at 15-minute intervals ($\Delta t = 0.25$) throughout the trading day. The error bars represent one standard deviation from the mean performance over the testing period. WamOL consistently achieves lower calibration and validation errors than the other models overall in both training and predictions. The performance of WamOL can be attributed to its effectiveness in incorporating market dynamics and no-arbitrage constraints into the calibration process.

\begin{figure}[htbp]
    \centering
    \includegraphics[width=1.0\columnwidth]{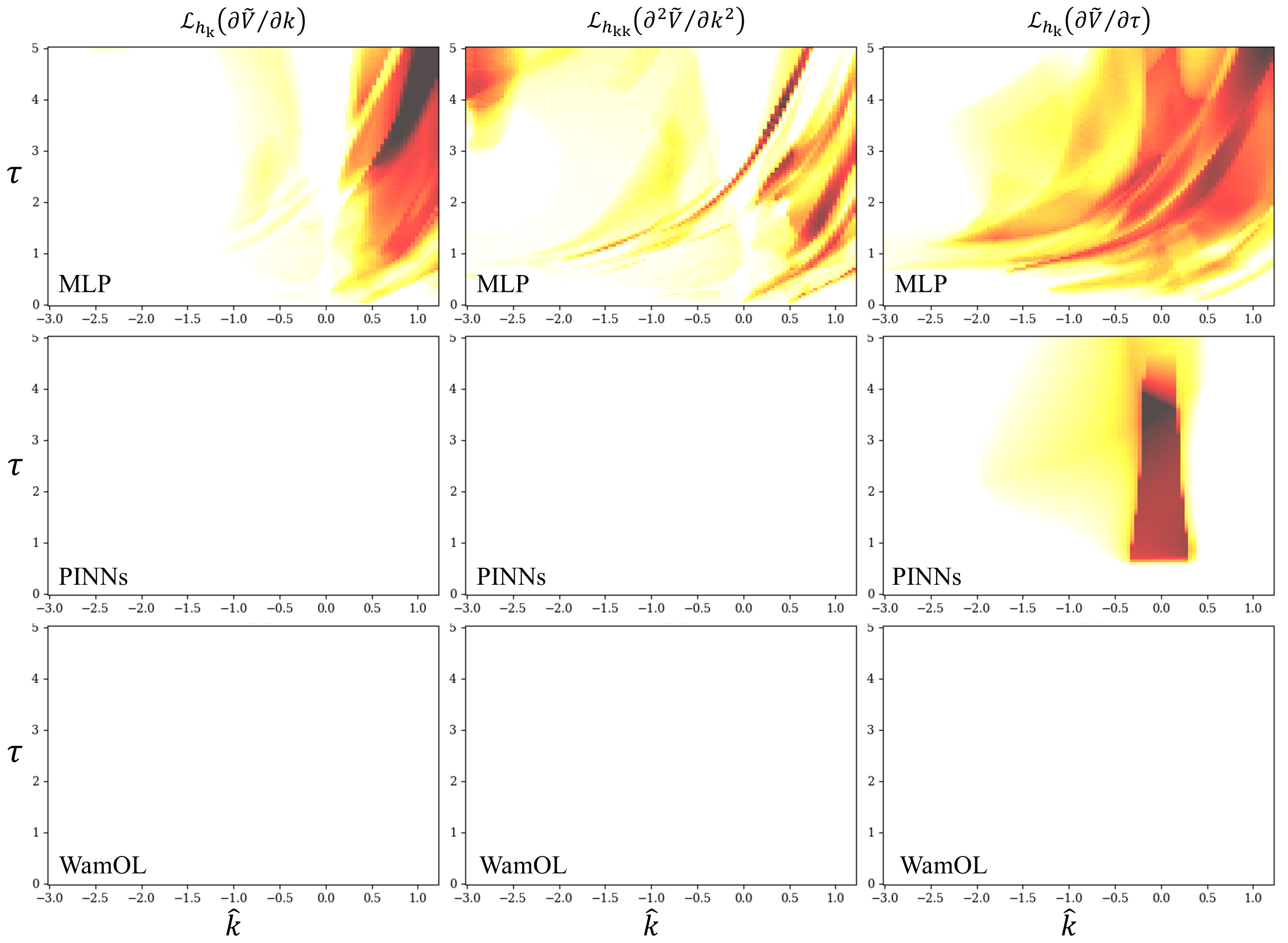}
    \caption{Heatmaps of no-arbitrage conditions of the IVS by models. Areas violating constraints are coloured red, with darker shades indicating higher degrees of violation, compressed over time dimension $t$.}
    \label{fig: wamol_arbitrage_heatmap}
    \Description[Heatmaps of the inequality derivative conditions for different models.]{Heatmaps of the inequality derivative conditions of the learned IVS for different models. Areas violating constraints are coloured red, with darker shades indicating higher degrees of violation compressed over the time dimension $t$.}
\end{figure}
In addition, we compare the performance of WamOL, which incorporates nonlinear constraints, by investigating the derivative profiles of the obtained surface. Figure \ref{fig: wamol_arbitrage_heatmap} illustrates heatmaps of the violations of the inequality derivative conditions, i.e., no-arbitrage conditions. The results demonstrate that the proposed WamOL framework captures the solution while satisfying the nonlinear constraints, outperforming the traditional PINNs approach.

These results highlight WamOL's ability to balance various objectives effectively. While it may not always achieve the lowest loss in each individual category during training, it demonstrates the best overall performance in the prediction phase. This indicates strong generalisation capabilities and a robust balance between fitting market prices and satisfying financial constraints. The MLP model, despite showing the best fit to market prices during training, performs poorly in prediction, suggesting overfitting. PINNs show improved performance over MLP but are generally outperformed by WamOL, especially in adhering to no-arbitrage constraints. Furthermore, the WamOL framework demonstrates versatility, as it can be applied to calibration problems involving design parameters in PDEs, including data-driven solutions in real-time market environments.

\subsection{Risk Sensitivity Analysis}
\label{subsection: Risk Sensitivity Analysis}
An important aspect of IVS calibration is the accurate estimation of risk sensitivities, such as delta, gamma, and theta. Delta measures the sensitivity of option prices to changes in the underlying asset price, gamma quantifies the rate of change of delta, and theta represents the sensitivity of option prices to changes in time to maturity. These risk sensitivities correspond to the inequalities in the no-arbitrage conditions, ensuring that the calibrated IVS adheres to fundamental financial principles and reflects market dynamics.

\begin{figure}[htbp]
    \centering
    \includegraphics[width=1.0\columnwidth]{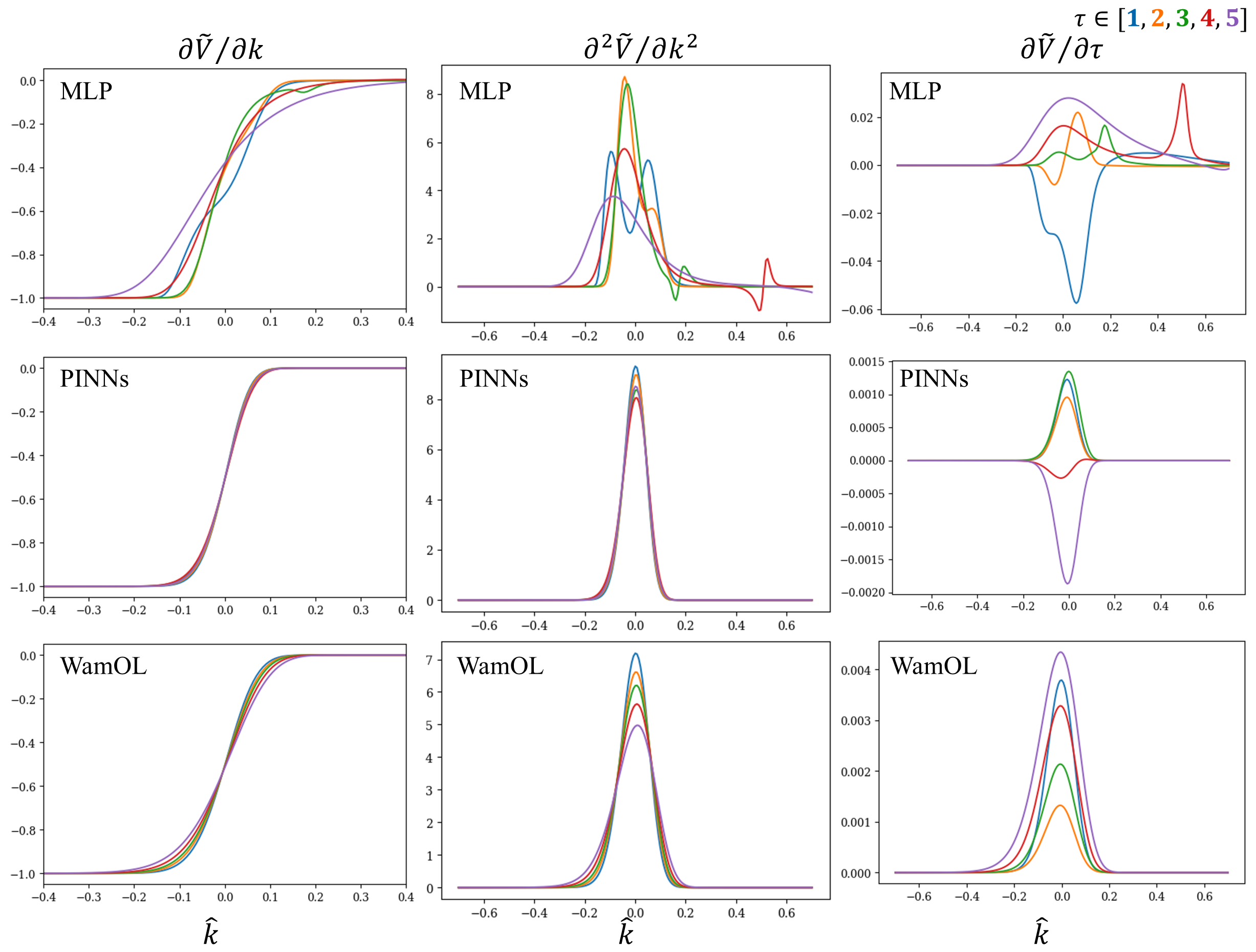}
    \caption{Risk sensitivity profiles associated with derivative profiles by models.}
    \Description[Risk sensitivity profiles generated by different models.]{Risk sensitivity profiles generated by different models. The figure illustrates the first and second derivatives with respect to log-moneyness widehat k (left and centre) and the first derivative with respect to time to maturity tau (right). Each line represents a slice of the surface at different maturities tau.}
    \label{figure: wamol_sens}
\end{figure}
Figure \ref{figure: wamol_sens} presents a comparative analysis of risk sensitivity profiles generated by WamOL and baseline models. WamOL demonstrates its ability to reasonably estimate variants of risk profiles, such as delta, gamma, and theta, while simultaneously satisfying no-arbitrage conditions and producing smooth, continuous risk sensitivity curves. In contrast, the baseline models exhibit notable discrepancies and irregularities, with the MLP model showing the most significant challenges in accurately capturing these sensitivities across different market scenarios.

The risk sensitivity comparisons underscore WamOL's superior capability to capture the underlying dynamics of the options market. This proficiency is particularly valuable for developing informed risk management and hedging strategies in option portfolio construction and effective management of risk exposure. The smooth and consistent profiles produced by WamOL provide practitioners with more reliable inputs for their financial decision-making processes, potentially leading to more robust and efficient risk management practices in complex market environments.

\subsection{Loss Scalability and Computational Efficiency}
\label{subsec:computational_efficiency_scalability}
\begin{figure}[htbp]
\centering
\includegraphics[width=\columnwidth]{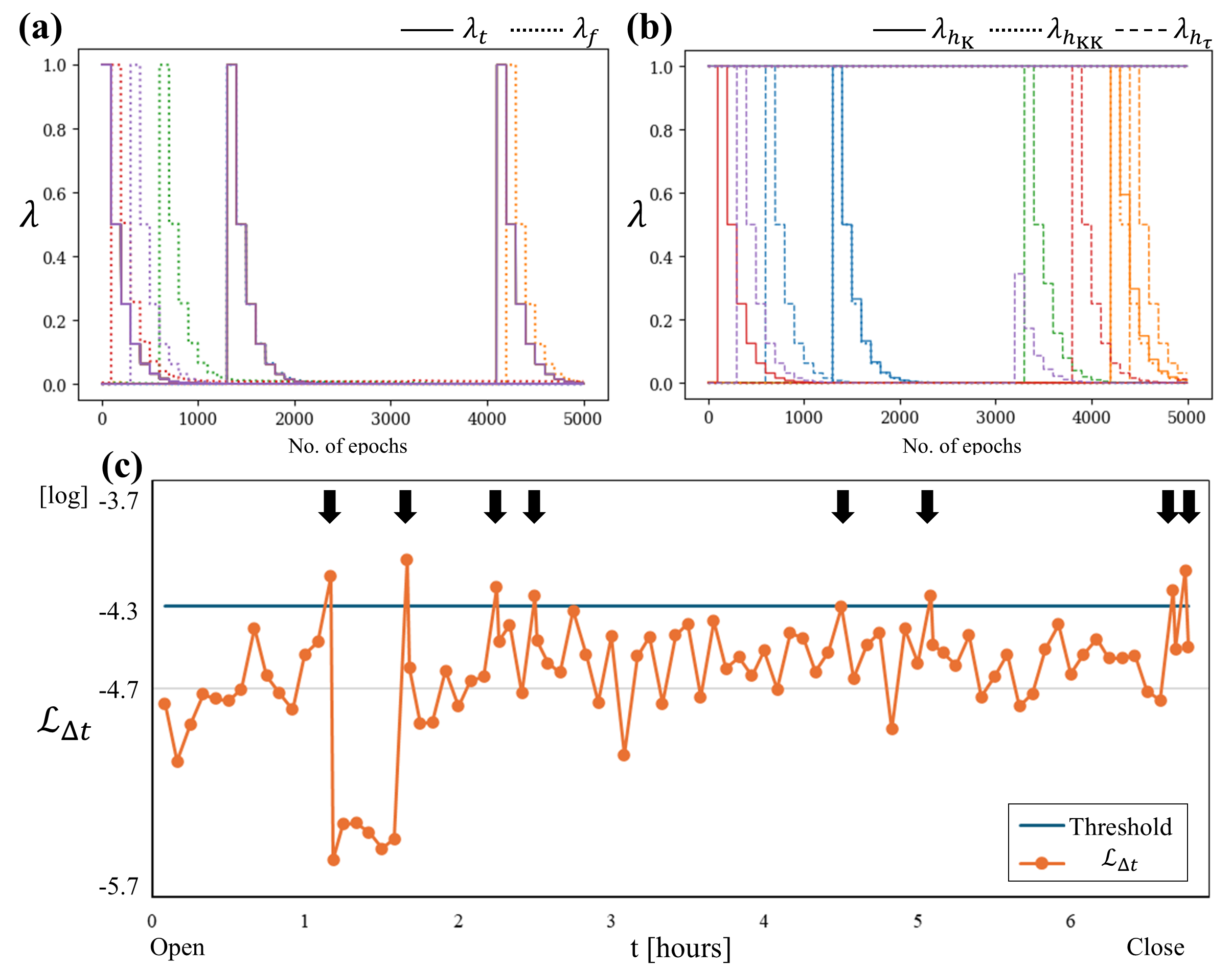}
\caption{\textbf{(a-b)} Evolution of loss weights for categorised losses in WamOL, normalised relative to the maximum value, across $5$ random training runs. \textbf{(c)} Intraday sample errors ($\Delta t \approx 0.083$) with online incremental learning, exemplified by data from 31st July 2023. Black arrows denote recalibration events.}
\label{fig: wamol_lambdas}
\Description[Loss weight evolution and observation errors in WamOL.]{Evolution of normalised loss weights during training and observation errors with online incremental learning in WamOL.}
\end{figure}
To demonstrate WamOL's effectiveness and computational efficiency, we analyse its behaviour during both training and real-time application. Figure \ref{fig: wamol_lambdas} (a-b) illustrates the evolution of normalised loss weights for each categorised loss across training epochs. Figure \ref{fig: wamol_lambdas} (c) depicts the observation errors during a typical trading day, showcasing the implementation of online incremental learning. The black arrows indicate recalibration events, triggered when prediction errors surpass a predefined threshold. 

The results demonstrate WamOL's capacity to react swiftly to derivative inequality violations, adapt to incoming data in real-time, and maintain prediction accuracy throughout the trading day. This dynamic parameter tuning, which resembles a "Whack-a-mole" game, shows its ability to adapt to the changing landscape of the optimisation problem, ensuring that the model converges to a solution that satisfies the derivative constraints while maintaining certain premium accuracy, useful in handling complex financial modelling scenarios.

\begin{table}[H]
    \centering
    \caption{The computation times (in seconds) for the training in Section \ref{subsection: Backtesting on Real Market Data} based on changes in dataset size ($N$) and number of neurons.}
    \label{tab: computational_efficiency}
    {\small\begin{tabular}{lccccc}
    \toprule
    \multirow{2}{*}{Models} & \multirow{2}{*}{Default} & \multicolumn{2}{c}{Dataset size} & \multicolumn{2}{c}{No. of neurons}\\
    && Half & Double & Half & Double \\
    \midrule
    MLP & $87.5$ & $73.2$ & $115.7$ & $63.4$ & $165.5$\\
    PINNs & $168.9$ & $151.6$ & $190.6$ & $104.6$ & $307.5$\\
    WamOL & $198.0$ & $180.7$ & $222.2$ & $132.1$ & $345.9$\\
    \bottomrule
    \end{tabular}}
\end{table}
Table \ref{tab: computational_efficiency} presents the computation times required by WamOL and the baseline models for calibrating the IVS on datasets of varying sizes. The computational efficiency of WamOL is evident from its ability to handle these complex optimisation challenges without significant overhead. The efficient use of automatic differentiation and the adaptive loss balancing approach contribute to WamOL's fast convergence and reduced computational overhead. WamOL also exhibits good scalability, as evidenced by the sublinear growth in computation time with respect to the dataset size. This scalability is crucial for handling the ever-increasing volumes of options data in modern financial markets, i.e., as real-time applications.

\balance
\section{CONCLUSIONS AND FUTURE WORK}
This study introduced Whack-a-mole Online Learning (WamOL), a novel calibration algorithm designed to address the multi-objective problem of derivatives, with a focus on Implied Volatility Surface (IVS) calibration in finance. WamOL demonstrated the ability to capture appropriate patterns from limited sparse data by integrating self-adaptive and auto-balancing processes, enabling reweighting mechanisms for each loss term.

Experiments conducted on frequent real-time market data highlight WamOL's effectiveness in accurately modelling the dynamics of the options market using PDEs while adhering to no-arbitrage constraints and minimising calibration errors, as well as improving intraday predictions. The adaptive loss balancing approach and the efficient use of automatic differentiation contributed to WamOL's fast convergence and reduced computational overhead. Its performance, adherence to PDEs and no-arbitrage constraints, and computational efficiency make WamOL a promising tool for practitioners in the financial industry as a real-time application.

Future research could focus on incorporating additional market information to enhance the realism of the model and better capture dependencies in options data. Although the soft constraints approach does not rigorously enforce each constraint in a multi-objective optimisation context, WamOL has demonstrated its ability to generate suitable surfaces when dealing with noisy real-world data and trade-off structures. While the experiments showcase the efficacy of the WamOL algorithm, further research could provide insights into how WamOL affects the loss landscape and explore its potential for application in other domains.

\begin{acks}
The authors are grateful to the Department of Computer Science, University College London, for providing us with the resources to perform this case study.
\end{acks}

\bibliographystyle{ACM-Reference-Format}
\bibliography{references}

\end{document}